\documentclass[manuscript,noblind]{geophysics}

\usepackage{array}
\usepackage{multirow}
\usepackage{tabularx}

\begin{document}
\title{Massive-scale unlabeled field and labeled synthetic seismic datasets of global shelf-edge clinothems}

\address{
\footnotemark[1] School of Earth and Space Sciences, State Key Laboratory of Precision Geodesy, Mengcheng National Geophysical Observatory, University of Science and Technology of China, Hefei 230026, China
\footnotemark[2]State Key Laboratory of Ocean Sensing \& Ocean College, Zhejiang University, Zhoushan, Zhejiang, 316021, China\\
\footnotemark[3] Institute of Advanced Technology, University of Science and Technology of China, Hefei, 230088, China\\
\footnotemark[4] School of Artificial Intelligence, State Key Laboratory of Digital Intelligent Technology for Unmanned Coal Mining, Anhui University of Science and Technology, Hefei 231131, China}
\author{Hui Gao\footnotemark[1], Xinming Wu\footnotemark[1], Jintao Li\footnotemark[2], Xiaoming Sun\footnotemark[3], and Jiarun Yang\footnotemark[4]}

\footer{Example}
\lefthead{Gao et al.}
\righthead{\emph{Geophysics}}
\renewcommand{\thefootnote}{\fnsymbol{footnote}} 
\renewcommand{\figdir}{figures} 

\begin{abstract}
Seismic stratigraphic interpretation of shelf-edge clinothems is essential for revealing tectonic evolution, paleoclimate change, depositional dynamic conditions, and hydrocarbon generation and accumulation during basin filling. However, traditional interpretation methods remain labor-intensive, time-consuming, and highly subjective. Although AI-based method offer a potential solution for automated this task, its development has been limited by the scarcity of comprehensive and representative benchmark datasets for shelf-edge clinothems. This limitation primarily arises from limited field data availability, the scarcity of reliable geological labels, and the structural complexity and strong variability of clinothem-dominated systems. To address this gap, we develop a hybrid benchmark dataset through two complementary strategies of field data curation and geological and geophysical forward modeling, ultimately generating 3,000 unlabeled field and 4,000 labeled synthetic seismic data, respectively. We further evaluate several representative baseline deep learning models on these datasets, and the accurate results demonstrate that the curated dataset provides an effective and representative basis for model training, quantitative assessment, and practical application. Finally, we have publicly released this hybrid benchmark dataset (https://doi.org/10.5281/zenodo.18910271) to facilitate the development, validation, and assessment of deep learning methods for automated seismic stratigraphic interpretation.
\end{abstract}

\section{Introduction}
Clinothem-dominated systems exhibit substantial variability in geometry, stacking pattern, and geological structures across different sedimentary basins~\citep{anell2017quantifiable,patruno2018clinoforms}, making manual interpretation difficult and posing a major challenge for developing robust and generalizable AI-based interpretation methods. Recently, several geophysical benchmark datasets, such as STEAD~\citep{mousavi2019stanford}, OpenFWI~\citep{deng2022openfwi}, cigChannels~\citep{wang2024cigchannel}, cigFacies~\citep{gao2025cigfacies}, OpenSWI~\citep{liu2025openswi} and GeoFWI~\citep{li2026geofwi}, have been developed to advance AI methods in various geophysical scenarios. However, such comprehensive and representative benchmark dataset remain highly scarce for shelf-edge clinothems of sedimentary basin. Moreover, most existing publicly available datasets of exploration geophysics primarily consist of synthetic data, while field datasets are typically limited in scale, lack geological labels, and are mainly utilized for testing and validation. Although synthetic datasets have proven effective for model training and have been successfully applied in many geophysical tasks~\citep{wu2020building,ferreira2024synthetic}, relying solely on synthetic data is insufficient for networks to fully capture the complexity and variability of field clinothems systems. This limitation is particularly pronounced in sedimentary basins, where numerous depositional hiatuses, erosional surfaces, unconformities, complex sequence architectures, and tectonic deformation make realistic forward modeling extremely challenging. Moreover, recent advances in self-supervised learning and large-scale pretraining have further increased the demand for diverse unlabeled field data, which provide realistic structural patterns and geological variability that are difficult to realistically simulate in synthetic data. In addition, field datasets spanning multiple basins are indispensable for evaluating model effectiveness, robustness, and generalization across diverse geological settings. Therefore, constructing a hybrid benchmark dataset that combines realistic field characteristics with high-quality geological labels is essential for effective modeling training, standardized baseline evaluation, and improved generalization in practical field applications.

To address these limitations, we construct a hybrid benchmark dataset consisting of an unlabeled field seismic dataset derived from field data curation and a labeled synthetic seismic dataset generated from geological and geophysical forward modeling (see Figure~\ref{fig:fig2_1}). Specifically, we curate publicly available field data through collection, manual selection, cropping, resampling, and standardization, ultimately building 3,000 unlabeled samples that preserve the realism, diversity, and structural complexity of clinothem-dominated systems. We then generate a labeled synthetic seismic dataset through a three-step workflow of stratigraphic forward modeling, geological structural deformation, and geophysical forward modeling, ultimately generating 4,000 geologically plausible synthetic samples with corresponding relative geologic time (RGT,~\cite{stark2003unwrapping,stark2004relative}) labels. These RGT labels provide a dense stratigraphic representation that preserves the full stratigraphic succession and relative temporal relationships, which is critical for characterizing clinothem architectures and supporting seismic stratigraphic interpretation. The field dataset provides realistic benchmark samples for field testing and application, whereas the synthetic dataset provides geologically reasonable and label-complete samples for supervised training and quantitative evaluation. Finally, we further evaluate several representative baseline models on these datasets to demonstrate their comprehensiveness, representativeness, geological plausibility, and effectiveness for model training and assessment.

\section{Benchmark Dataset Construction}
Relative geologic time (RGT) estimation is a continuous regression task that is highly sensitive to stratigraphic continuity and structural consistency. Its performance depends not only on network architecture and optimization strategy, but also critically on the representativeness of the training data and the quality of the target labels. In this study, we develop two complementary strategies to construct massive-scale training datasets with diverse clinothems architectures shown in Figure~\ref{fig:fig2_1}. The first strategy is to build 3,000 unlabeled field seismic data through field data curation (Figure~\ref{fig:fig2_1}a). This field dataset preserves the realism, diversity, structural complexity and representativeness of field clinothem-dominated systems, thereby  providing an essential basis for model evaluation and field application. The second strategy is to build 4,000 labeled synthetic seismic data through geological and geophysical forward modeling (Figure~\ref{fig:fig2_1}b). This synthetic dataset provides geologically consistent samples with corresponding RGT labels, enabling the model to learn a stable mapping from seismic data to RGT.

\plot{fig2_1}{width=\textwidth}
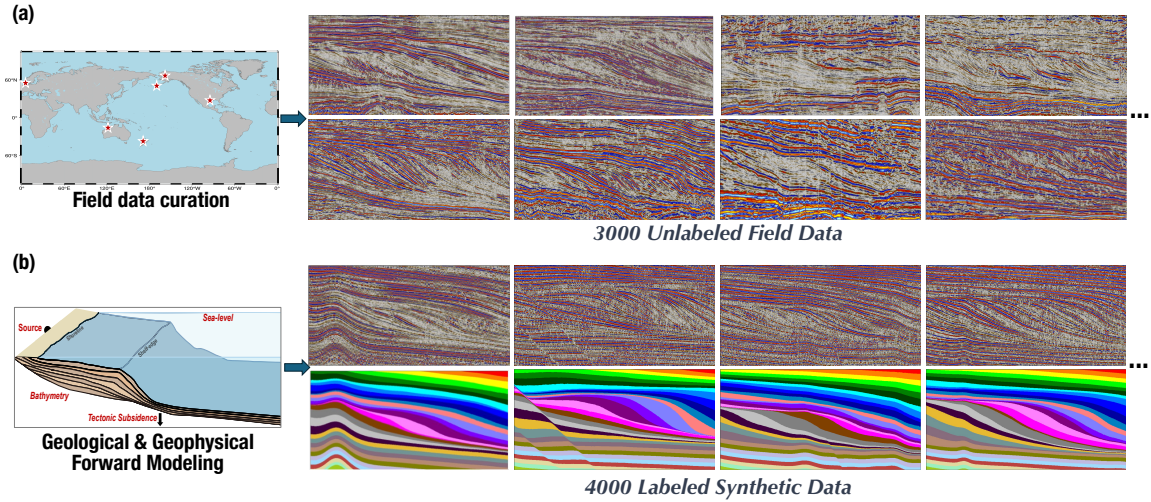
{Construction of the unlabeled field and labeled synthetic seismic datasets. (a) We construct the unlabeled field dataset through 2D and 3D raw data collection, manual selection, cropping, resampling, and data post-processing, resulting in 3,000 normalized 2D field seismic images with diverse clinothems architectures. (b) We then perform stratigraphic forward modeling, geological structural deformation, and geophysical forward modeling to generate various realistic synthetic seismic datasets and corresponding RGT labels.}


\subsection{Unlabeled Field Seismic Datasets}
To construct a large-scale field dataset containing diverse clinothem architectures across different sedimentary basins, we develop a systematic field data curation workflow consisting of raw data collection, manual selection and cropping, resampling, and data post-processing. Specifically, we first collect more than 600 3D seismic volumes and 2,000 2D seismic profiles from multiple publicly available dataset repositories, such as the Society of Exploration Geophysicists (SEG), Netherlands Oil and Gas Portal (NLOG), New Zealand Petroleum \& Minerals (NZPAM), the United States Geological Survey (USGS), the South Australian Resources Information Gateway (SARIG), and Western Australian Petroleum and Geothermal Information Management System (WAPIMS). We then manually select and crop these datasets to retain approximately 60 3D seismic volumes and 500 2D seismic profiles that exhibit representative clinothems geometries. These selected field data are primarily distributed across the Dutch North Sea, Australia, New Zealand, Alaska North Slope, the Beaufort Sea, and Gulf of Mexico, encompassing diverse sedimentary environments and stratigraphic patterns (see Figure~\ref{fig:fig2_2}). We further randomly slice the 3D volumes into 2D profiles from inline or crossline directions with at least 25 profiles between adjacent slices to reduce redundancy. Finally, we perform resampling and mean–variance normalization process to each profile, yielding 3,000 unlabeled 2D field seismic data, as illustrated in Figures~\ref{fig:fig2_1}a and \ref{fig:fig2_2}.

\plot{fig2_2}{width=\textwidth}
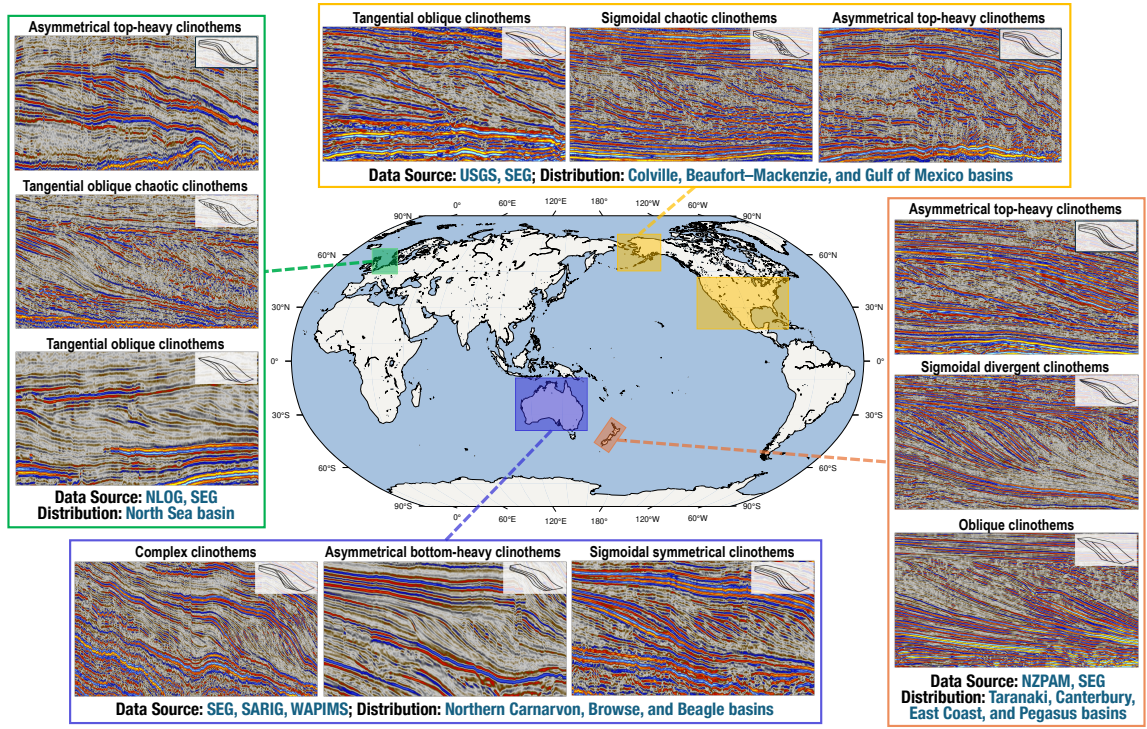
{Spatial distribution and representative patterns of field seismic clinothem dataset. These field samples covers several representative sedimentary basins worldwide and contains diverse clinothems geometric types. These geometric cartoon diagrams are modified from \cite{anell2017quantifiable}. }


\tabl{table1}{A summary of the field and synthetic seismic datasets.
}{
  \begin{center}
    \begin{tabular}{m{3.2cm}|m{7.1cm}|m{1.6cm}|m{1cm}}
      \hline
      Dataset & Category & Number & Total \\
      \hline
      \multirow{7}{*}{Field Dataset}
      & North Sea Basin & 1982 & \multirow{7}{*}{3000} \\
      \cline{2-3}
      & New Zealand East Coast, Raukumara, Pegasus, Taranaki, Canterbury Basins & 596 \\
      \cline{2-3}
      & Australia Northern Carnarvon, Browse, Beagle Basins & 180 \\
      \cline{2-3}
      & Gulf of Mexico, Colville, Beaufort–Mackenzie Basins & 242 \\
      \hline
      \multirow{4}{*}{Synthetic Dataset}
      & One 3rd-order sea-level cycle & 780 & \multirow{4}{*}{4000}  \\
      \cline{2-3}
      & Two 3rd-order sea-level cycles & 1370 \\
      \cline{2-3}
      & Three 3rd-order sea-level cycles & 1714 \\
      \cline{2-3}
      & Four 3rd-order sea-level cycles & 136 \\
      \hline
    \end{tabular}
  \end{center}
}

The field clinothem dataset provides broad coverage of sedimentary settings and clinothem geometries, highlighting its diversity, complexity, and representativeness. In terms of sedimentary environments, such datasets covers multiple representative basin surveys, including North Sea, Northern Carnarvon, Browse, Beagle, New Zealand East Coast, Raukumara, Pegasus, Taranaki, Canterbury, Gulf of Mexico, Colville, Beaufort–Mackenzie basins (Table~\ref{tbl:table1}), spanning a wide range of sedimentary dynamic conditions, long-term tectonic and climate changes, and depositional environments. In terms of clinothem geometric complexity, these datasets contains a variety of typical clinothem architectures (Figure~\ref{fig:fig2_2}), such as oblique, tangential oblique, tangential oblique chaotic, asymmetric top-heavy, asymmetric bottom-heavy, sigmoidal symmetric, sigmoidal divergent, sigmoidal chaotic, complex clinothems types \citep{anell2017quantifiable}. These geometric variations reflect diverse stratigraphic stacking patterns formed under different conditions of sediment supply, accommodation, tectonic processes, and climate changes. Although this dataset does not contain RGT labels, it preserves the geometric variability and realistic stratigraphic and structural features, thus providing a valuable and comprehensive benchmark and application dataset for assessing model performance across different sedimentary basin. Besides, this unlabeled field dataset can be used for unsupervised or self-supervised learning, enabling networks to learn representative field stratigraphic and structural features from diverse geological settings, thus potentially reducing the domain gap between synthetic and field data.

\subsection{Labeled Synthetic Seismic Datasets}
\plot{fig2_3}{width=\textwidth}
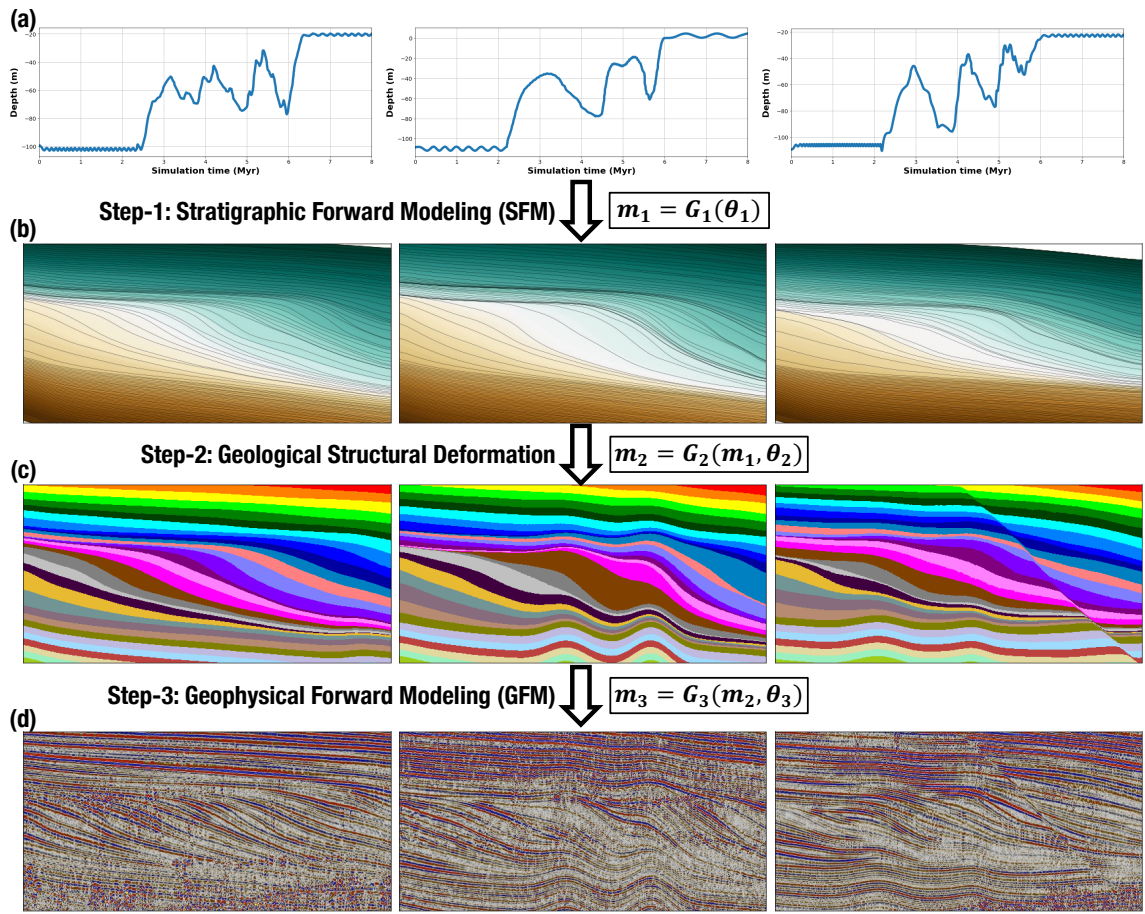
{Geological and geophysical forward modeling workflow for constructing labeled synthetic seismic dataset. (a) Relative sea level curves, (b) stratigraphic layers simulating from stratigraphic forward modeling, (c) geological models after geological structural deformation, and (d) synthetic seismic images generated from geophysical forward modeling.}


To address the scarcity of high-quality RGT labels in field seismic data, we further develop a comprehensive and geologically plausible three-step forward modeling workflow modified from \cite{gao2025geologically} to generate realistic synthetic seismic data and corresponding labels. This workflow consists of stratigraphic forward modeling, geological structural deformation, and geophysical forward modeling. We first simulate sediment transport and depositional evolution in fluvial–deltaic systems using sedimentary process simulation, and generate various shelf-edge stratigraphic clinoforms model ($\mathbf{m}_1$) with realistic and diverse stratigraphic stacking patterns and complex geometric architectures shown in Figure~\ref{fig:fig2_3}b. This process can be expressed as:
\begin{equation}
\mathbf{m}_1 = \mathbf{G}_1(\mathbf{\theta}_1),\\
\label{eq1}
\end{equation}
where $\mathbf{G}_1$ represents the stratigraphic forward modeling operator implemented with a landscape evolution model \citep{salles2018pybadlands}, and $\mathbf{\theta}_1$ denotes a set of geologically reasonable forcing conditions, such as initial bathymetry, river discharge and distribution, sediment flux, sea-level fluctuations (Figure~\ref{fig:fig2_3}a and Table~\ref{tbl:table1}), tectonic subsidence, and climate changes. To further improve the realism and diversity of geological models under complex post-depositional geological structural deformations, we then add faulting and folding structures into stratigraphic models ($\mathbf{m}_1$) to simulate realistic subsurface geological models ($\mathbf{m}_2$) during post-depositional burial (Figure~\ref{fig:fig2_3}c). This deformation process is formulated as: 
\begin{equation}
\mathbf{m}_2 = \mathbf{G}_2(\mathbf{m}_1, \mathbf{\theta}_2),\\
\label{eq2}
\end{equation}
where $\mathbf{G}_2$ represents the operator of simulating geological structural deformation proposed by \cite{wu2020building}, and $\mathbf{\theta}_2$ denotes a set of parameters controlling fault and fold geometries, such as fault spatial distributions, strikes, dips, and throws, as well as fold locations, extents, and shapes. To simulate realistic fault characteristics within field shelf-edge regions, we restrict its strike to primarily subparallel with the basin margin, thus aligning with the typical normal fault systems commonly developed in extensional tectonic environments. Finally, we perform geophysical forward modeling ($\mathbf{G}_3$) to generate the corresponding realistic synthetic seismic data ($\mathbf{d}$), which is given by:
\begin{equation}
\mathbf{d} = \mathbf{G}_3(\mathbf{m}_2, \mathbf{\theta}_3),\\
\label{eq3}
\end{equation}
where $\mathbf{G}_3$ represents geophysical forward modeling operator, including realistic impedance model construction, depth-time conversion, wavelet convolution, and real or random noise addition, and $\mathbf{\theta}_3$ denotes a set of parameters controlling the geophysical properties, wavelet type and dominant frequency, noise characteristics, and signal-to-noise ratio. Through geological and geophysical forward modeling workflow, we ultimately generate 4,000 synthetic seismic data and corresponding RGT labels, as shown in Figures~\ref{fig:fig2_1}b and \ref{fig:fig2_3}d. Such synthetic dataset maintains both geological plausibility and physical consistency while providing accurate and high-quality RGT labels, thereby establishing a reliable supervision foundation for AI models to learn a stable seismic-to-RGT mapping.

\section{Baseline Model Evaluations}

\plot{fig2_4}{width=\textwidth}
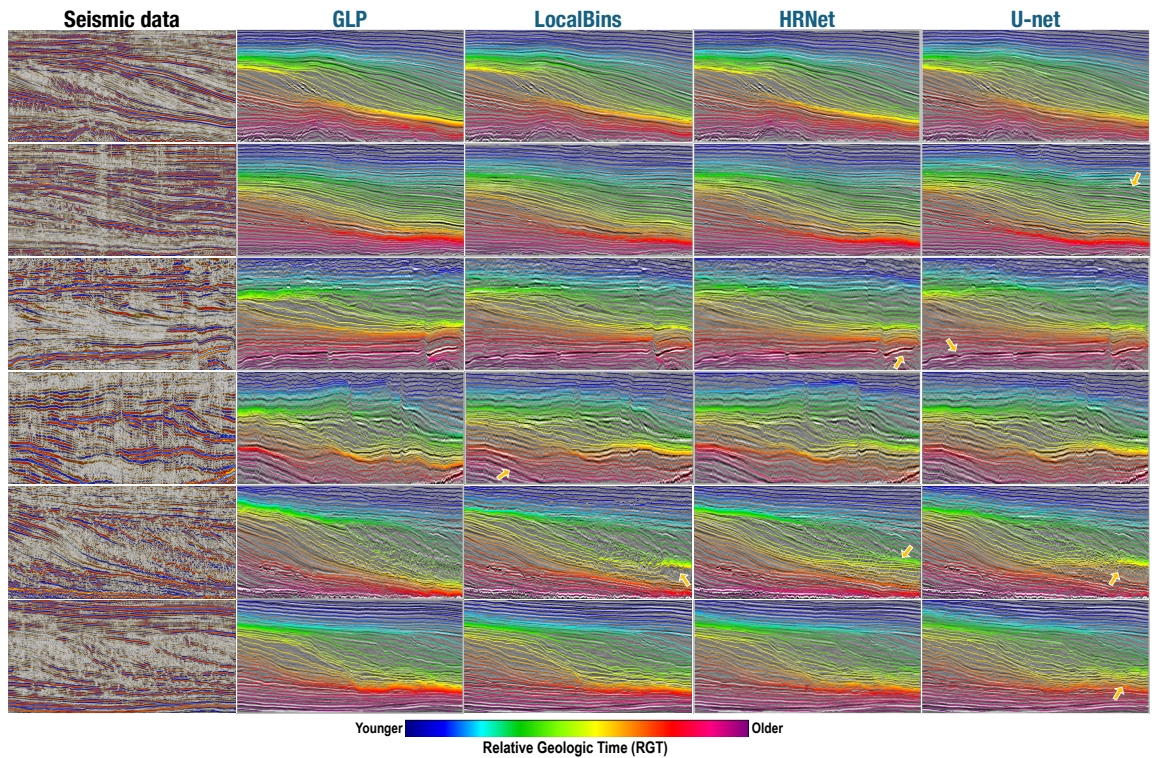
{Comparison of RGT predicted results across different baseline models (GLP, LocalBins, HRNet, and U-Net). Multiple iso-surfaces are extracted from the RGT predictions and overlaid on the seismic image.}


To evaluate the effectiveness of the curated benchmark datasets, we further perform a comprehensive performance evaluation using several representative deep learning models, including U-Net~\citep{ronneberger2015u}, High-Resolution Network (HRNet, \cite{wang2020deep}), LocalBins~\citep{bhat2022localbins}, and Global-Local Path Networks (GLP, \cite{kim2022global}). These baseline models are trained and evaluated within a geologically informed and data-driven framework modified from \cite{gao2025geologically}, which integrates synthetic supervision and geological prior constraints for training. We then randomly select six samples from field dataset and apply the trained baseline models for RGT estimation, and extract multiple iso-surfaces at equal geologic time intervals, as illustrated in Figure~\ref{fig:fig2_4}. The extracted iso-surfaces exhibit strongly alignment with the seismic reflectors and provide a reliable representation of clinothems geometries, stratigraphic stacking patterns, geological structues, and unconformities distribution across various field surveys. These precise results demonstrate that the curated benchmark dataset provides an effective and representative basis for model training, practical application, and quantitative evaluation. However, some fine-scale weak stratigraphic fittings are still observed in local regions with stratigraphic pinch-out, unconformities, low signal-to-noise ratios, and complex tectonic structures, as indicated by the orange arrows in Figure~\ref{fig:fig2_4}. Besides, the performance comparison reveals that transformer-based networks (GLP and LocalBins) and multi-scale feature-fusion network (HRNet) outperform the simpler convolutional U-Net. This performance gap highlights the critical importance of extracting and sensing global features for RGT estimation in shelf-edge clinothem data, which commonly contain numerous depositional hiatuses and stratigraphic discontinuities.

\section{CONCLUSION}
In this paper, we construct a comprehensive hybrid benchmark dataset specifically designed for shelf-edge clinothems, comprising 3,000 unlabeled field seismic samples and 4,000 labeled synthetic seismic samples with corresponding RGT labels. Baseline model evaluations demonstrate that these two complementary benchmark datasets provide an effective and representative foundation for model training, field application, quantitative evaluation, development and deployment of AI method for seismic stratigraphic interpretation in clinothem-dominated systems. Finally, we have made these comprehensive clinothem datasets and representative deep learning baseline models publicly available at https://doi.org/10.5281/zenodo.18910271~\citep{gao2026dataset}, facilitating further model development, comparison, and validation for high-resolution seismic stratigraphic interpretation of sedimentary basins.

\section{ACKNOWLEDGMENTS}
The authors thank SEG, NLOG, NZPAM, USGS, SARIG, and WAPIMS for providing public available datasets. We also thank the USTC supercomputing center for providing computational resources.


\section{Data and Materials Availability}
The comprehensive labeled synthetic and unlabeled field datasets of global shelf-edge clinothems are publicly available in Zenodo~\citep{gao2026dataset}, and the corresponding source code has been released on GitHub for further research.

\bibliographystyle{plainnat} 
\bibliography{huig26data}

\end{document}